\documentclass[fleqn,10pt]{wlscirep}
\usepackage{graphicx, amsmath, amsfonts, amssymb}
\usepackage{authblk}
\usepackage{float}

\title{Two-dimensional Dirac particles in a P{\"o}schl-Teller waveguide}

\author[1,*,+]{R. R. Hartmann}
\author[2,$\dagger$,+]{M. E. Portnoi}
\affil[1]{
Physics Department,
De La Salle University,
2401 Taft Avenue,
Manila 1004,
Philippines.}
\affil[2]{
School of Physics, University of Exeter, Stocker Road, Exeter EX4 4QL, United Kingdom and International Institute of Physics, Universidade Federal do Rio Grande do Norte, Natal - RN, Brazil
}
\date{}

\affil[*]{richard.hartmann@dlsu.edu.ph}
\affil[$\dagger$]{m.e.portnoi@exeter.ac.uk}

\affil[+]{these authors contributed equally to this work}

\begin{abstract}
We obtain exact solutions to the two-dimensional (2D) Dirac equation for the one-dimensional P{\"o}schl-Teller potential which contains an asymmetry term. The
eigenfunctions are expressed in terms of Heun confluent functions,
while the eigenvalues are determined via the solutions of a simple
transcendental equation. For the symmetric case, the eigenfunctions
of the supercritical states are expressed as spheroidal wave functions, and approximate analytical expressions are obtained for the corresponding eigenvalues. 
A universal condition for any square integrable symmetric potential is obtained for the minimum strength of the potential required to hold a bound state of zero energy. Applications for smooth electron waveguides in 2D Dirac-Weyl systems are discussed.
\end{abstract}

\begin{document}
\flushbottom
\maketitle
\thispagestyle{empty}
\noindent

\section*{Introduction}
The P{\"o}schl-Teller potential \cite{poschl1933bemerkungen}
plays an important role in many fields of physics \cite{dong2007factorization} from modeling diatomic molecules and quantum many-body systems \cite{RomerPRL93,RomerPRB94a,RomerPRB94b}, to applications in astrophysics \cite{ferrari1984new, berti2009quasinormal}, optical waveguides \cite{kogelnik19752} and quantum wells \cite{radovanovic2000intersubband,yildirim2005nonlinear} through to Bose-Einstein and Fermionic condensates \cite{baizakov2004delocalizing,antezza2007dark}, and supersymmetric quantum mechanics \cite{dutt1988supersymmetry}. For the one-dimensional Schr{\"o}dinger equation, the hyperbolic
symmetric form can be solved in terms of associated Legendre polynomials
and the eigenvalues are known explicitly \cite{poschl1933bemerkungen,landau2013quantum}.
We consider an analogous relativistic problem, that of a two-dimensional Dirac particle, confined by a one-dimensional P{\"o}schl-Teller
potential. Several solutions have been obtained for the Dirac
equation with central P{\"o}schl-Teller potentials \cite{jia2007solutions,xu2008approximate,wei2009spin,wei2009algebraic,jia2009approximate,miranda2010solution,tacskin2011spin} and the hyperbolic-secant potential is also known to admit analytic solutions for both the one and two-body one-dimensional Dirac problems \cite{hartmann2010smooth,hartmann2014quasi,hartmann2011excitons}. Modified P{\"o}schl-Teller potential potentials have also been employed in numerical simulations of potential barriers in bilayer graphene \cite{park2015two}.

With the recent explosion of research in Dirac materials \cite{wehling2014dirac}
there has been a renewed interest in quasi-relativistic phenomena considered in condensed matter systems of different dimensionalities. This is due to the fact that the same equations which govern Dirac fermions in relativity, map directly to the equations of motion
describing the quasi-particles in systems such as graphene \cite{neto2009electronic},
carbon nanotubes \cite{charlier2007electronic}, topological insulators \cite{hasan2010colloquium,qi2011topological,PhysRevB.94.014502}, transition
metal dichalcogenides \cite{xiao2012coupled} and 3D Weyl semimetals \cite{young2012dirac}. Massless Dirac particles are notoriously difficult to confine; however, it has been demonstrated that certain types of one-dimensional electrostatic
waveguides in graphene, possess zero energy-modes which are truly
confined within the waveguide \cite{hartmann2010smooth,hartmann2014quasi} and that the number
of these zero energy-modes is equal to the number of supercritical
states (i.e. bound states whose energy, $E=-M$, where $M$ is
the particle's effective mass). Transmission resonances and supercriticality of Dirac particles through one-dimensional potentials have been studied extensively \cite{Coulter_AJP_71,Calogeracos_PN_96,PhysRevA.58.2160,Dombey_PRL_20,Kennedy_JPA_02,Villalba_PRA_03,Kennedy_IJMPA_04,Guo_EJP_09,Villalba_PS_10,Sogut_PS_11,Arda_PS_11,miserev2016analytical}. The majority of studies model top-gated carbon-based nanostructures using abrupt potentials \cite{katsnelson2006chiral,Chaplik_06,PhysRevB.74.045424,Chau_PRB_09,Zhang_APL_09,Williams_NanoT_11,Yuan_JAP_11,Wu_APL_11,ping2012oscillating,Katsnelson_PS_12,PhysRevB.94.165443}.
However, experimental potential profiles vary smoothly over many lattice constants, with even the smallest of gate generated potentials being far from square \cite{FoglerPRL}. There is also some controversy concerning waveguides which are defined by smoothly decaying, square integrable functions, which decay at large distances as $1/x^{n}$, where $n>1$. Numerical experiments imply that such  waveguides always contain a zero-energy mode \cite{stone2012searching}, whereas analysis based upon relativistic Levinson theorem, says there is a minimum potential strength required to observe a zero-energy mode \cite{clemence1989low,lin1999levinson,calogeracos2004strong,ma2006levinson,miserev2016analytical}. Our result supports the latter, and we demonstrate this through a simple analysis of supercritical states of zero energy.

In 2D Dirac materials, the low-energy spectrum of the
charge carriers can be described by a Dirac Hamiltonian \cite{dirac1928quantum}
of the form
\begin{equation}
\hat{H}=\hbar v\left(\sigma_{x}\hat{k}_{x}+s_{\mathrm{K}}\sigma_{y}\hat{k}_{y}+\sigma_{z}k_{z}\right),\label{eq:Ham_orig}
\end{equation}
where $\hat{k}_{x}=-i\frac{\partial}{\partial x}$, $\hat{k}_{y}=-i\frac{\partial}{\partial y}$,
$\sigma_{x,y,z}$ are the Pauli spin matrices. $v$ plays the role
of the speed of light and $k_{z}$ is proportional to the particle in-plane effective mass. In graphene, the charge carriers are massless, $k_{z}=0$, and
the dispersion is linear, $v=v_{\mathrm{F}}\approx10^{6}$~m/s is
the Fermi velocity and $s_{\mathrm{K}}$ has the value of $\pm1$
for the $K$ and $K'$ valley respectively \cite{wallace1947band}. For narrow-gap nanotubes and certain types of graphene nanoribbons \cite{dutta2010novel,chung2016electronic},
the operator $\hat{k}_{y}$ can by replaced by the number $k_{y}=E_{g}/(2 \hbar v_{\mathrm{F}} )$, which in the absence of applied field is fixed by geometry, where $E_{g}$ is the value of the bandgap. For nanotubes, $E_{g}$ can be controlled by applying a magnetic field along the nanotube axis \cite{portnoi2008terahertz,hartmannoptoelectronic,hartmann2011excitons,hartmann2014terahertz,hartmann2015terahertz}. When $k_{z}$ is finite, Eq.~(\ref{eq:Ham_orig})
can be used as a simple model for silicene \cite{liu2011low} or Weyl semimetal \cite{wehling2014dirac,hills2017current}. It has also been proposed that when graphene is subjected
to a periodic potential on the lattice scale, for example, graphene
on top of a lattice-matched hexagonal boron nitride \cite{giovannetti2007substrate}
the Dirac fermions acquire mass with $E_{g}=2\hbar v_{\mathrm{F}}k_{z}$
being of the order of 53 meV.

In what follows we shall consider a particle described by the Hamiltonian (\ref{eq:Ham_orig}) subjected
to a one-dimensional potential $U\left(x\right)$, which varies on
a scale much larger than the lattice constant of the corresponding Dirac material, therefore allowing us to neglect inter-valley scattering for the case of graphene. We shall also set $s_{\mathrm{K}}=1$,
and note that the other valley's wave function can be obtained by
exchanging $k_{y}\rightarrow-k_{y}.$ The Hamiltonian acts on the
two-component Dirac wavefunction
\begin{equation}
\Psi=e^{ik_{y}y}\left(\begin{array}{c}
\Psi_{A}\left(x\right)\\
\Psi_{B}\left(x\right)
\end{array}\right)
\end{equation}
to yield the coupled first-order differential equations
\begin{equation}
\left(V-E+M\right)\Psi_{A}-i\left(\frac{d}{d\tilde{x}}+\Delta\right)\Psi_{B}=0
\label{eq:ham_1}
\end{equation}
and
\begin{equation}
\left(V-E-M\right)\Psi_{B}-i\left(\frac{d}{d\tilde{x}}-\Delta\right)\Psi_{A}=0.
\label{eq:ham_2}
\end{equation}
where $\tilde{x}=x/L$ and $L$ is a constant. $V=UL/\hbar v_{\mathrm{F}}$
and the charge carrier energy, $\varepsilon$, have been scaled such
that $E=\varepsilon L/\hbar v_{\mathrm{F}}$. The charge carriers
propagate along the $y$-direction with wave vector $k_{y}=\Delta/L$
, which is measured relative to the Dirac point, $\Psi_{A}\left(x\right)$
and $\Psi_{B}\left(x\right)$ are the wavefunctions associated with the
$A$ and $B$ sublattices of graphene and finally $M=k_{z}L$ represents
an effective mass in dimensionless units.

In what follows we consider the relativistic quasi-one-dimensional P{\"o}schl-Teller
potential problem which can be applied to describe e.g. graphene waveguides. We obtain the
exact energy eigenfunctions for this potential and formulate a method
for calculating the eigenvalues of the bound states. We then analyze
the energy-spectrum of the symmetric P{\"o}schl-Teller potential
and obtain expressions for the eigenvalues of the supercritical states. By analyzing the zero-energy supercritical states we obtain a universal threshold condition for the minimum potential strength required for a potential to possess a zero-energy mode, for any square-integrable potential. We also show that the eigenfunctions in the non-relativistic limit restore the one-dimensional Schr{\"o}dinger equation solutions. Finally, we analyze the eigenvalue spectrum for the modified P{\"o}schl-Teller potential which includes an asymmetry term.

\section*{Relativistic one-dimensional P{\"o}schl-Teller problem}
\begin{figure}[ht]
    \centering
    \includegraphics[viewport=0 0 480 360,width=0.5\textwidth]{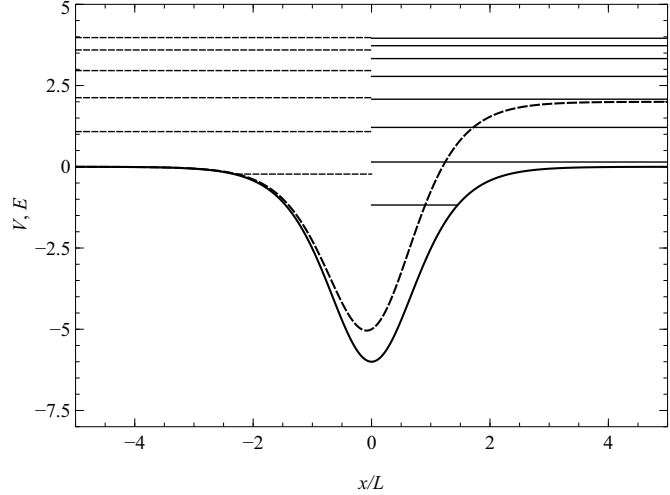}
    \caption{
The solid line shows the modified P{\"o}schl-Teller potential, Eq.~\ref{eq:potential}, for the symmetric case of $a=24$ and $b=0$. The dashed line shows the asymmetric potential for the case of $a=24$ and $b=2$. The 8 solid horizontal lines are the bound state energy levels for the symmetric potential at $\Delta = 4$ and the 6 dashed horizontal lines are the bound state energy levels for the asymmetric potential at $\Delta = 4$.
}
\label{fig:eigenval_both}
\end{figure}

In this section we consider the potential
\begin{equation}
V=-\frac{a}{4}\left[1-\tanh^{2}\left(
\tilde{x}
\right)\right]+\frac{b}{2}\left[1+\tanh\left(
\tilde{x}
\right)\right],
\label{eq:potential}
\end{equation}
which is a linear combination of the symmetric P{\"o}schl-Teller potential with
an additional term which enables the introduction of asymmetry \cite{nieto1978exact}. This potential belongs to the class of quantum models, which are quasi-exactly solvable \cite{turbiner1988quantum,ushveridze1994quasi,bender1998quasi,downing2013solution,hartmann2014bound,hartmann2014quasi,HartmannAIP,PhysRevA.95.062110}, where
only some of the eigenfunctions and eigenvalues are found explicitly.
The depth of the well is given by $-\left(a-b\right)^{2}/4a$, and the potential width is characterized by the parameter $L$, which was introduced after Eq.~(\ref{eq:ham_2}). For the case of $b=0$, the potential transforms into the symmetric P{\"o}schl-Teller potential,
while if $a=0$, the potential is a smooth potential step, which can
be used to model a p-n junction \cite{miserev2012quantum,Katsnelson_AoP_13}. The symmetric
and asymmetric forms of the potential are plotted in Fig.~\ref{fig:eigenval_both}.

Substituting $\Psi_{A}=\left(\Psi_{1}+\Psi_{2}\right)/2$ and $\Psi_{B}=\left(\Psi_{1}-\Psi_{2}\right)/2$
allows Eqs.~(\ref{eq:ham_1}-\ref{eq:ham_2}) to be reduced to a
single second-order differential equation in $\Psi_{1}$ $\left(\Psi_{2}\right)$
\begin{equation}
\left[\left(V-E\right)^{2}-M^{2}-\Delta^{2}+is\frac{dV}{d\tilde{x}}\right]\Psi_{j}+\frac{d^{2}\Psi_{j}}{d\tilde{x}^{2}}=0,
\label{eq:second_order_org}
\end{equation}
where $s=-\left(-1\right)^{j}$
and
$j=1,\,2$ correspond to the spinor components
$\Psi_{1}$ and $\Psi_{2}$ respectively. By making the natural change of variable $Z=\left[1+\tanh\left(\tilde{x}\right)\right]/2$
and using the wave function of the form $\Psi_{j}=\exp\left(pZ\right)Z^{n}\left(1-Z\right)^{m}\psi_{j}\left(Z\right)$
allows Eq.~(\ref{eq:second_order_org}) to be reduced to the Heun
confluent equation \cite{heun1888theorie} in
variable $Z$:
\begin{equation}
\frac{\partial^2 \psi_{j}}{\partial Z^{2}}+\frac{\alpha Z^{2}+\left(2-\alpha+\beta+\gamma\right)Z-1-\beta}{Z\left(Z-1\right)}\frac{\partial \psi_{j}}{\partial Z}
+\frac{\left[\left(2+\beta+\gamma\right)\alpha+2\delta_{s}\right]Z-\left(1+\beta\right)\alpha+\left(1+\gamma\right)\beta+\gamma+2\eta_{s}}{2Z\left(Z-1\right)}\psi_{j}=0,
\label{eq:HeunEq.}
\end{equation}
where $n=\beta/2$, $m=\gamma/2$, $P=\alpha/2$, $\widetilde{\Delta}^{2}=M^{2}+\Delta^{2}$, $s_{\alpha,\,\beta,\,\gamma}=\pm1$
and the parameters $\alpha,\,\beta,\,\gamma,\,\delta_{s}$ and $\eta_{s}$
are:
$\alpha=ias_{\alpha}$, $\beta=s_{\beta}\sqrt{\widetilde{\Delta}^{2}-E^{2}}$, $\gamma=s_{\gamma}\sqrt{\widetilde{\Delta}^{2}-\left(E-b\right)^{2}}$, $\delta_{s}=a\left(\frac{1}{2}b-is\right)$ and $\eta_{s}=\beta^{2}/2-\left(a-b\right)\left(E-is\right)/2$.
This same method of reducing a system of coupled first-order differential equations to the Heun confluent equation has been exploited to solve various generalisations of the quantum Rabi model \cite{XieJPA17}, and notably the quasi-exact solutions of the P{\"o}schl-Teller family potentials and Rabi systems are closely related \cite{RamazanJPA02}. In some instances, the resulting Heun confluent functions can be terminated as a finite polynomial \cite{PhysRevLett.57.787, hartmann2014quasi} allowing particular eigenvalues to be obtained exactly, providing the parameters obey special relations. When this is not the case, the energy spectrum can be obtained fully via the Wronskian method \cite{ZhongJPA2013, hartmann2014quasi,MACIEJEWSKI201416, XieJPA17}, which is the method we shall utilize.

Equation~(\ref{eq:HeunEq.}) has regular singularities at $Z=0$ and $1$,
and an irregular one at $Z=\infty$ which is outside the domain of
$\tilde{x}$. The solutions to Eq.~(\ref{eq:HeunEq.}) are given by
\begin{equation}
\psi_{j}=\sum_{s_{\alpha},\,s_{\beta},\,s_{\gamma}}A_{j,\,s_{\alpha},\,s_{\beta},\,s_{\gamma}}H\left(\alpha,\,\beta,\,\gamma,\,\delta_{s}\,,\eta_{s};\,Z\right)
+
B_{j,\,s_{\alpha},\,s_{\beta},\,s_{\gamma}}Z^{-\beta}H\left(\alpha,\,-\beta,\,\gamma,\,\delta_{s}\,,\eta_{s};\,\,Z\right),
\end{equation}
where $A_{j,\,s_{\alpha},\,s_{\beta},\,s_{\gamma}}$ and $B_{j,\,s_{\alpha},\,s_{\beta},\,s_{\gamma}}$ are constants, and $H$ is the Heun confluent function \cite{ronveaux1995heun}, which
has a value of $1$ at the origin. For $\left|Z\right|<1,$ $H\left(\alpha,\,\beta,\,\gamma,\,\delta_{s}\,,\eta_{s},\,Z\right)=\left(1-Z\right)^{-\gamma}H\left(\alpha,\,\beta,\,-\gamma,\,\delta_{s}\,,\eta_{s},\,Z\right)$
and $H\left(\alpha,\,\beta,\,\gamma,\,\delta_{s}\,,\eta_{s},\,Z\right)=\exp\left(-\alpha Z\right)H\left(-\alpha,\,\beta,\,\gamma,\,\delta_{s}\,,\eta_{s},\,Z\right)$, therefore, as expected from the general theory of second order differential equations, the full solution is given by a linear combination of just two linearly independent functions
\begin{eqnarray}
\Psi_{j}=
\left[
A_{j}
H\left(\alpha,\,\beta,\,\gamma,\,\delta_{s}\,,\eta_{s},\,Z\right)Z^{\frac{1}{2}\beta}
+
B_{j}
H\left(\alpha,\,-\beta,\,\gamma,\,\delta_{s}\,,\eta_{s},\,Z\right)Z^{-\frac{1}{2}\beta}
\right]
\left(1-Z\right)^{\frac{1}{2}\gamma}
\exp\left(\frac{1}{2}\alpha Z\right),
\label{eq:Full_Left}
\end{eqnarray}
where $\beta$ is not an integer. It should be noted that when $a=0$
and $\alpha=\delta=0$, the Heun confluent functions appearing in Eq.~(\ref{eq:Full_Left})
reduces to a Gauss hypergeometric functions and for the case of a massless
particle Eq.~(\ref{eq:Full_Left}) reduce down to the solutions obtained in Ref. \cite{miserev2012quantum}. If, however, $\beta$ is an integer then $H\left(\alpha,\,-\beta,\,\gamma,\,\delta\,,\eta_{s},\,Z\right)$
is divergent and $B_{i}$ has to be set to zero, and the second linearly independent
solution can be constructed from a series expansion about the point
$1-Z$. The solutions to the Heun confluent equation thus
far have been given as power series expansions about the point $Z=0$.
However, these power series rapidly diverge as $Z$ approaches the
second singularity; therefore, at $Z=1$ we must construct solutions
as power series expansions in variable $1-Z$:
\begin{equation}
\Psi_{j}=
\left[
C_{j}H\left(-\alpha,\,\gamma,\,\beta,\,-\delta_{s},\,\eta_{s}+\delta;\,1-Z\right)\left(1-Z\right)^{\frac{1}{2}\gamma}
+
D_{j}H\left(-\alpha,\,-\gamma,\,\beta,\,-\delta_{s},\,\eta_{s}+\delta;\,1-Z\right)\left(1-Z\right)^{-\frac{1}{2}\gamma}
\right]
Z^{\frac{1}{2}\beta}\exp\left(\frac{1}{2}\alpha Z\right).
\label{Full_Right}
\end{equation}
The constants $C_{i}$ and $D_{i}$ are found by matching the two power series
expansions and their derivatives at $Z_{0}$ where $0<Z_{0}<1$.

For bound states, we require that $\widetilde{\Delta}^{2}>E^{2}$
and $\widetilde{\Delta}^{2}>\left(E-b\right)^{2}$. These conditions ensure that the bound states are inside the effective bandgap (which accounts for the motion along the y-axis). As $x\rightarrow-\infty$,
$Z\rightarrow0$ and as $x\rightarrow\infty$, $Z\rightarrow1$ ,
therefore we may write the asymptotic expressions of $\Psi_{j}$ as
\begin{equation}
\lim_{Z\rightarrow0}\left(\Psi_{j}\right)=
A_{j}Z^{\frac{1}{2}\beta}+B_{j}Z^{-\frac{1}{2}\beta}
\end{equation}
and
\begin{equation}
\lim_{Z\rightarrow1}\left(\Psi_{j}\right)=\left[C_{j}\left(1-Z\right)^{\frac{1}{2}\gamma}+D_{j}\left(1-Z\right)^{-\frac{1}{2}\gamma}\right]\exp\left(\frac{1}{2}\alpha\right).
\end{equation}
Therefore, for bound states, $B_{i}$ ($A_{i}$) is zero for $s_{\beta}=1$
($s_{\beta}=-1$) and $D_{i}$ ( $C_{i}$) is zero for $s_{\gamma}=1$
($s_{\gamma}=-1$). Clearly the choice of $s_{\beta}$ and $s_{\gamma}$ is
arbitrary, therefore, from hereon in we set both to $1$ unless otherwise
stated. In this instance, the energy eigenvalues are found from the condition:
\begin{equation}
\left.\frac{\partial H\left(\alpha,\,\beta,\,\gamma,\,\delta_{s}\,,\eta_{s},\,Z\right)}{\partial Z}\right|_{Z_{0}}
H\left(-\alpha,\,\gamma,\,\beta,\,-\delta_{s},\,\eta_{s}+\delta;\,1-Z_{0}\right)
=
\left.\frac{\partial H\left(-\alpha,\,\gamma,\,\beta,\,-\delta_{s},\,\eta_{s}+\delta;\,1-Z\right)}{\partial Z}\right|_{Z_{0}}
H\left(\alpha,\,\beta,\,\gamma,\,\delta_{s}\,,\eta_{s},\,Z_{0}\right),
\label{eq:bound_state_condition}
\end{equation}
where $0<Z_{0}<1$.
\subsection*{Symmetric P{\"o}schl-Teller potential solutions}
In general, relating $\Psi_{1}$ to $\Psi_{2}$ is non-trivial,
since neither a known expression exists which connects Heun confluent functions
about two different singular points for arbitrary parameters, nor
is there a general expression relating the derivative of the confluent
Heun function to other confluent Heun functions, though particular
instances have been obtained \cite{fiziev2009novel,ShahnazaryanNR}.
However, for the symmetric P{\"o}schl-Teller potential (i.e. $b=0$) one can obtain the relation:
\begin{eqnarray}
2Z\left(1-Z\right)\frac{dH\left(\alpha,\,\beta,\,\gamma,\,\delta_{1}\,,\eta_{1},\,Z\right)}{dZ}&=&
\left(\beta-iE\right)H\left(\alpha,\,\beta,\,\gamma,\,\delta_{-1}\,,\eta_{-1},\,Z\right)
\nonumber
\\
&+&
\left[\left(\beta+\gamma\right)Z-\left(\beta-iE\right)-\left(\alpha+\delta_{1}\right)Z\left(1-Z\right)\right]H\left(\alpha,\,\beta,\,\gamma,\,\delta_{1}\,,\eta_{1},\,Z\right).
\label{eq:deriv_ident}
\end{eqnarray}
Therefore, $A_{2}=\Omega_{s_{\beta}}A_{1}$ and $B_{2}=\Omega_{-s_{\beta}}B_{1}$, where $\Omega_{s_{\beta}}=\left(E+i\beta\right)/\left(M+i\Delta\right)$.

In pristine graphene, only the symmetric form of Eq.~(\ref{eq:potential})
will contain non-leaky modes at zero energy. Non-zero-energy modes will have a finite lifetime since they can always couple to continuum states outside of the waveguide, whereas zero-energy modes are fully confined since the density of states vanishes at zero energy outside of the well. Asymmetric forms of Eq.~(\ref{eq:potential}) never contain truly bound modes since the density of states cannot vanish on both sides of the potential simultaneously. Notably, this is somewhat counterintuitive as for the Schr{\"o}dinger problem a symmetric potential always contains a bound state, which can be removed by asymmetry. The emergence of bound states for a relativistic problem with an infinitely wide barrier is a manifestation of the Klein tunneling phenomenon~\cite{hartmann2010smooth,hartmann2014quasi}.

We shall now consider the symmetric form of Eq.~(\ref{eq:potential}) for massless particles. Accordingly, we set $b=0$ and $M=0$, and in this instance the symmetrized real functions~\cite{hartmann2010smooth,hartmann2014quasi} are given by $\Psi_{\mathrm{I}}=\Psi_{A}+i\Psi_{B}$
and $\Psi_{\mathrm{II}}=\Psi_{A}-i\Psi_{B}$, where
\begin{equation}
\Psi_{A}=A_{1}\Re\left[\Phi\exp\left(\frac{1}{2}\alpha Z-i\frac{\theta}{2}\right)
\right]
Z^{\frac{1}{2}\beta}\left(1-Z\right)^{\frac{1}{2}\gamma}\exp\left(i\frac{\theta}{2}\right)
\label{eq:Psi_A_prof}
\end{equation}
and
\begin{equation}
\Psi_{B}=iA_{1}\Im\left[\Phi\exp\left(\frac{1}{2}\alpha Z-i\frac{\theta}{2}\right)
\right]
Z^{\frac{1}{2}\beta}\left(1-Z\right)^{\frac{1}{2}\gamma}\exp\left(i\frac{\theta}{2}\right),
\label{eq:Psi_B_prof}
\end{equation}
where $\Phi=H\left(\alpha,\,\beta,\,\gamma,\,\delta_{1}\,,\eta_{1},\,Z\right)$ and $\tan\theta=-E/\beta$. By employing the identity Eq.(\ref{eq:deriv_ident}), the derivatives appearing in Eq.(\ref{eq:bound_state_condition}) can be expressed in terms of Heun confluent functions. It immediately follows that at the origin $\Psi_{\mathrm{I}}^{\star}\Psi_{\mathrm{II}}+\Psi_{\mathrm{I}}\Psi_{\mathrm{II}}^{\star}=0$, which in terms of the functions $\Psi_{A}$ and $\Psi_{B}$ yields:
\begin{equation}
\left|\Psi_{A}\left(Z=\frac{1}{2}\right)\right|^{2}=\left|\Psi_{B}\left(Z=\frac{1}{2}\right)\right|^{2},
\label{eq:eigen_eq_abv}
\end{equation}
where $Z=1/2$ corresponds to $x=0$. Substituting Eq.(\ref{eq:Psi_A_prof}) and Eq.(\ref{eq:Psi_B_prof}) into Eq.(\ref{eq:eigen_eq_abv}) results in the condition
\begin{equation}
\Re\left[\Phi\left(Z=\frac{1}{2}\right)\exp\left(\frac{\alpha}{4}-i\frac{\theta}{2}\right)\right]\mp
\Im\left[\Phi\left(Z=\frac{1}{2}\right)\exp\left(\frac{\alpha}{4}-i\frac{\theta}{2}\right)\right]=0.
\label{eq:bound_state_fast}
\end{equation}
From Eq.(\ref{eq:Psi_A_prof}) and Eq.(\ref{eq:Psi_B_prof}), $A_{1}\Psi_{A}^{\star}=A_{1}^{\star}\Psi_{A}\exp\left(-i\theta\right)$ and $A_{1}\Psi_{B}^{\star}=-A_{1}^{\star}\Psi_{B}\exp\left(-i\theta\right)$. Therefore, the condition from which the eigenvalues of the spectrum are determined, Eq.(\ref{eq:eigen_eq_abv}), can be written as $\left.\left(\Psi_{A}+i\Psi_{B}\right)\right|_{Z=\frac{1}{2}}\left.\left(\Psi_{A}-i\Psi_{B}\right)\right|_{Z=\frac{1}{2}}=0$. This condition can be understood in terms of parity. In principle, one can construct from these functions odd and even solutions. However, since the even modes of $\Psi_{\mathrm{I}}$ occur at the same energies as the odd modes of $\Psi_{\mathrm{II}}$ and vice versa, one can obtain the eigenvalues when the symmetrized functions $\Psi_{\mathrm{I}}$ or $\Psi_{\mathrm{II}}$ are zero at the origin.
The functions $\Psi_{\mathrm{I}}$ and $\Psi_{\mathrm{II}}$ are related to the earlier introduced functions $\Psi_{1}$ and $\Psi_{2}$ by
\begin{equation}
\Psi_{\mathrm{I}}=\frac{1}{2}\left[\left(1+i\right)\Psi_{1}+\left(1-i\right)\Psi_{2}\right]
\label{eq:connection_A}
\end{equation}
and
\begin{equation}
\Psi_{\mathrm{II}}=\frac{1}{2}\left[\left(1-i\right)\Psi_{1}+\left(1+i\right)\Psi_{2}\right].
\label{eq:connection_B}
\end{equation}
Using Eqs. (\ref{eq:connection_A},\ref{eq:connection_B}) together with Eqs. (\ref{eq:Full_Left},\ref{Full_Right}) allows $\Psi_{\mathrm{I}}$ and $\Psi_{\mathrm{II}}$ to be expressed explicitly in terms of Heun functions.

\begin{figure}[ht]
    \centering
    \includegraphics[viewport=0 0 420 420,width=0.5\textwidth]{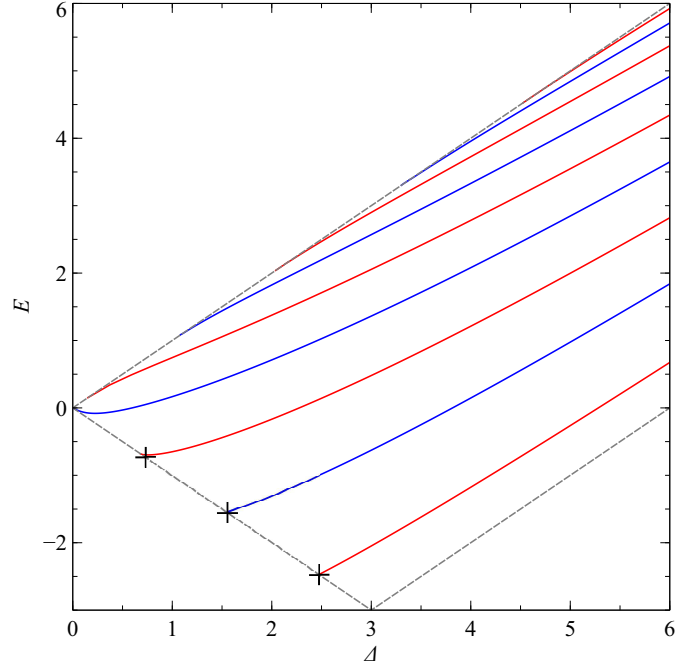}
    \caption{
(Color online) The energy spectrum of confined states in the symmetric P{\"o}schl-Teller potential, of strength $a=24$, as a function of $\Delta$. The alternating red and blue lines represent the odd (even) and even (odd) modes of $\Psi_{\mathrm{I}}$ ($\Psi_{\mathrm{II}}$) respectively. The black crosses denote the supercritical states. The boundary at which the bound states merge with the continuum is denoted by the grey (short-dashed) lines.
    }
    \label{Spectrum_b0_approx}
\end{figure}

\begin{figure}[ht]
    \centering
    \includegraphics[width=0.5\textwidth]{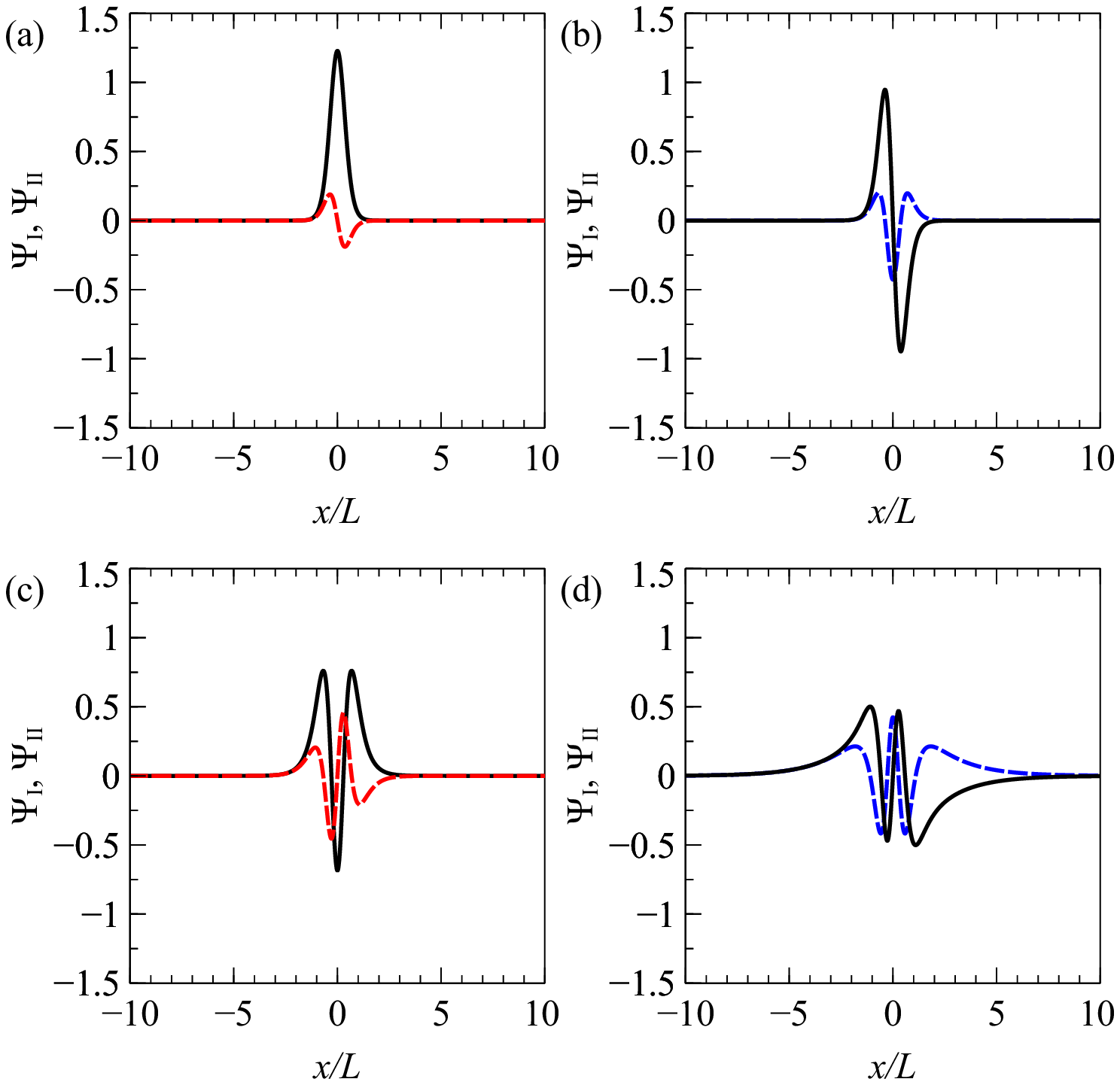}
\caption{
(Color online) The normalized zero-energy bound-state wavefunctions of the symmetric P{\"o}schl-Teller potential, for strength $a=24$. (a), (b), (c) and (d) correspond to the case of  $\Delta = 0.597,\,2.276,\, 3.817$ and $5.282$, respectively. The dashed-red and dashed-blue lines correspond to $\Psi_{\mathrm{I}}$, while the solid-black lines correspond to $\Psi_{\mathrm{II}}$.
}
\label{fig:eigenfunctions_zero}
\end{figure}

Eq.~(\ref{eq:bound_state_fast}) is formally the same as
Eq.~(\ref{eq:bound_state_condition}) but computationally faster. In Fig.~\ref{Spectrum_b0_approx} we plot the numerically obtained solutions of Eq. (\ref{eq:bound_state_fast}) for the potential defined by $a=24$ and $b=0$. The dashed lines represent the boundary at which the bound states merge with the continuum which occurs at the energies $E=\pm\Delta$ and $E+a/4=\Delta$. For the potential of strength $a=24$ we find that there are four zero-energy bound modes, occurring at $\Delta = 0.597, \,2.276, \,3.817$ and $5.282$. Their normalized wavefunctions are shown in Fig.~\ref{fig:eigenfunctions_zero}.


As mentioned previously, the number of zero-energy modes is equal
to the number of supercritical states (neglecting the spin and valley degrees of freedom). For the symmetric P{\"o}schl-Teller potential, the eigenvalues of these supercritical states can be determined approximately, via simple analytic expressions.
Moving to the symmetric basis, $\left(\Psi_{\mathrm{I}},\,\Psi_{\mathrm{II}}\right)^{\mathrm{T}}$, allows the pair of coupled first order differential equations, Eq.~(\ref{eq:ham_1}) and Eq.~(\ref{eq:ham_2}) to be reduced to a single second order differential equation in $\Psi_{\mathrm{II}}$:
\begin{equation}
\left[\left(V-E\right)^{2}-\Delta^{2}\right]\Psi_{\mathrm{II}}-\frac{1}{\left(V-E-\Delta\right)}\frac{dV}{d\tilde{x}}\frac{d\Psi_{\mathrm{II}}}{d\tilde{x}}+\frac{d^{2}\Psi_{\mathrm{II}}}{d\tilde{x}^{2}}=0.
\label{eq:sym_equation}
\end{equation}
For supercritical states, $E=-\Delta$, Eq.~(\ref{eq:sym_equation}) transforms into the differential equation for the angular prolate spheroidal wave functions~\cite{abramowitz1964handbook}:
\begin{equation}
\frac{d}{d\eta}\left[\left(1-\eta^{2}\right)\frac{d}{d\eta}S_{1N}\left(c,\eta\right)\right]+\left[\lambda_{1N}-c^{2}\eta^{2}-\frac{1}{1-\eta^{2}}\right]S_{1N}\left(c,\eta\right)=0,
\end{equation}
where $\eta=\tanh\left(z\right)$, $c=\pm V_{0}$, $\Psi_{\mathrm{II}}=\sqrt{1-\eta^{2}}S_{1N}$, and $S_{1N}$ are the spheroidal
wave functions. $\lambda_{1N}=a\left(a-8\Delta\right)/16$, where the permissible values of $\lambda_{1N}$ must be determined to assure that $S_{1N}\left(c,\eta\right)$ are finite at
$\eta=\pm1$. The permissible $\lambda_{1N}$ can be obtained via the asymptotic expansion
\begin{equation}
\lambda_{1N}\left(c\right)=cq+1-\frac{1}{8}\left(q^{2}+5\right)-\frac{q}{64c}\left(q^{2}+11-32\right)+O\left(c^{-2}\right),
\label{eq:First_bound_expand}
\end{equation}
where $N=1,\,2,\,3,\,\ldots$ and $q=2N-1$~\cite{abramowitz1964handbook}. Keeping only the terms of expansion shown in Eq.~(\ref{eq:First_bound_expand}) yields the following eigenvalues:
\begin{equation}
E=-\frac{1}{2}\left(1-2N\right)-\frac{a}{8}+\frac{1}{4a}\left[3-\left(1-2N\right)^{2}\right],
\label{eq:First_bound_guess}
\end{equation}
where $N$ is restricted to ensure that $E$ is negative. The resulting approximate eigenvalues for the symmetric P{\"o}schl-Teller potential of strength $a=24$ are $E=-2.475$, $-1.555$ and $-0.734$ respectively, and are indicated as black crosses in Fig.~\ref{Spectrum_b0_approx}. The approximate eigenvalues deviate increasingly from the numerically exact results, $E=-2.473$, $-1.542$ and $-0.682$, with decreasing $y$. It should be noted that a refinement of the approximate values of $\lambda_{1N}$ can be found in Ref.~\cite{abramowitz1964handbook}.

For the hyperbolic secant potential, $V=-V_{0}/\cosh\left(\tilde{x}\right)$, it was found that there was a minimum potential strength of $V_{0}=1/2$, required to observe a zero-energy mode \cite{hartmann2010smooth}. According to the Landauer formula, the conductance along the waveguide when the Fermi level is set to the Dirac point is $4ne^{2}/h$, where $n$ is the number of zero-energy modes. The existence of a threshold in the potential strength needed for the waveguide to contain a zero-energy mode allowed us to suggest that such waveguides could be used as switchable devices. However, later numerical calculations utilizing a variable phase method implied that power-decaying potentials always possess a bound mode \cite{stone2012searching}. This result cast serious doubt in the validity of employing exponentially decaying potentials as a suitable model for graphene waveguides, since realistic potential profiles decay a power of distance rather than exponentially.
Notably, the threshold potential strength at which the first zero-energy mode appears can be obtained from the condition of the first bound state coinciding with the
first supercritical state, i.e. $E=-\Delta=0$. In this instance, Eq.~(\ref{eq:sym_equation}) can be solved exactly:
\begin{equation}
\Psi_{\mathrm{I}}=C_{1}\cos\left(\int_{0}^{\tilde{x}}V\left(X\right)dX\right)+C_{2}\sin\left(\int_{0}^{\tilde{x}}V\left(X\right)dX\right)
\label{eq:Thres_1},
\end{equation}
\begin{equation}
\Psi_{\mathrm{II}}=-C_{1}\sin\left(\int_{0}^{\tilde{x}}V\left(X\right)dX\right)+C_{2}\cos\left(\int_{0}^{\tilde{x}}V\left(X\right)dX\right)
\label{eq:Thres_2}.
\end{equation}
For even modes of $\Psi_{\mathrm{I}}$, $C_{2}=0$, whereas, for odd modes of $\Psi_{\mathrm{I}}$, $C_{1}=0$. In the absence of the potential, when $E=\Delta=0$ the two first order differential equations in $\Psi_{\mathrm{I}}$ and $\Psi_{\mathrm{II}}$ decouple, and Eq. (\ref{eq:sym_equation}) reduces to a first order differential equation. As $E=-\Delta\rightarrow0$ and $x\rightarrow\pm\infty$ (where the potential is zero), Eq. (\ref{eq:Thres_1}) and Eq. (\ref{eq:Thres_2}) are required to be linearly dependent \cite{DombeyJPA02} and the Wronskian of the solutions $\Psi_{\mathrm{I}}$ and $\Psi_{\mathrm{II}}$ is zero\cite{calogero1967variable}. Consequently, the bound modes satisfy the condition: $\left|\Psi_{\mathrm{I}}\left(\pm\infty\right)\right|^{2}=\left|\Psi_{\mathrm{II}}\left(\pm\infty\right)\right|^{2}$. Therefore, the threshold potential strength at which the first zero-energy mode appears is found by the condition
\begin{equation}
\left|\int_{0}^{\infty}V\left(\tilde{x}\right)d\tilde{x}\right|=\frac{\pi}{4}.
\label{eq_threshold}
\end{equation}
Therefore, for any square-integrable potential, the threshold for the appearance of the first bound state of zero energy is only a function of the integrated potential. Notably, this is the same result obtained as relativistic Levinson theorem \cite{clemence1989low,lin1999levinson,calogeracos2004strong,ma2006levinson,miserev2016analytical}.
For the P{\"o}schl-Teller potential, Eq.~(\ref{eq_threshold}) yields $a=\pi$, which agrees with Eq.~(\ref{eq:bound_state_fast}). For $V=-V_{0}/\cosh\left(\tilde{x}\right)$, Eq.~(\ref{eq_threshold}) yields $V_{0}=1/2$ which restores the result of Ref. \cite{hartmann2010smooth}. This result implies that square-integrable power decaying potentials do indeed have a threshold, in contrast to the numerically predicted result of \cite{stone2012searching}. In this respect, exponentially decaying potentials are not that different from power-decaying and are perfectly suitable for the modeling of top-gated Weyl semimetals.

Finally, it should be noted that in the non-relativistic limit, Eq.~(\ref{Full_Right})
restores the well known results~\cite{landau2013quantum} for the bound state functions of the Schr{\"o}dinger equation for the P{\"o}schl-Teller potential. In the limit that $\alpha$
and $\delta$ are much smaller than $\beta$ and $\gamma$ (i.e. large $\Delta$):
\begin{equation}
\lim_{\Delta\rightarrow\infty}\left(H\left(\alpha,\,\beta,\,\gamma,\,\delta\,,\eta,\,Z\right)\right)=
Z^{-\beta}\,_{2}F_{1}\left(Q,\,1+\gamma-\beta-Q;\,1+\gamma;\,1-Z\right),
\end{equation}
where
$Q=\left(1+\gamma-\beta\pm\sqrt{1+\gamma^{2}+\beta^{2}-4\left(\eta_{s}+\delta_{s}\right)}\right)/2$
and $\,_{2}F_{1}$ is the Gauss hypergeometric function. Substituting
$E=E_{\mathrm{SE}}+\Delta$, $b=0$ and $s_{\beta}=-s_{\gamma}=1$ into Eq.~(\ref{Full_Right}),
results in the non-relativistic bound state functions
\begin{equation}
\lim_{\Delta\rightarrow\infty}\left(\Psi_{1}\right)\propto\,_{2}F_{1}\left(\epsilon+1+T,\,\epsilon-T;\,1+\epsilon;\,1-Z\right)Z^{\frac{\epsilon}{2}}\left(1-Z\right)^{\frac{\epsilon}{2}},
\end{equation}
where $\epsilon=s_{\beta}\sqrt{-2E_{\mathrm{SE}}\Delta}$
and $T=\left(-1+\sqrt{1+2a\Delta}\right)/2$. For the solutions to be finite at $Z=0$, we require that $\epsilon-T=-N$ where $N=0,\,1,\,2,\,\ldots$. When this criteria is met, the
Gauss hypergeometric function is a polynomial of degree $N$ and the
energy levels are given by
\begin{equation}
E_{\mathrm{SE}}=-\frac{1}{8\Delta}\left[-\left(1+2N\right)+\sqrt{1+2a\Delta}\right]^{2}.
\end{equation}

\subsection*{Modified P{\"o}schl-Teller potential solutions}
\begin{figure}[ht]
    \centering
    \includegraphics[width=0.75\textwidth]{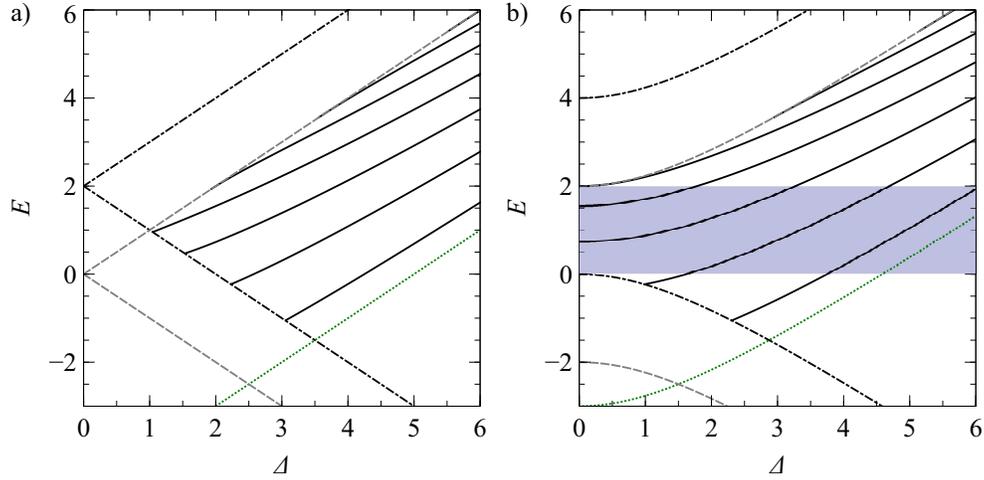}
    \caption{
(Color online) The energy spectra of confined states in the modified P{\"o}schl-Teller potential, of strength $a=24$ and $b=2$, as a function of $\Delta$ for (a) $M=0$ and (b) $M=2$. The grey (dashed), black (dot-dashed) and green (dotted) lines correspond to $\Delta^{2}+M^{2}=E^{2}$, $\Delta^{2}+M^{2}=\left(E-b\right)^{2}$ and $\Delta^{2}+M^{2}=\left[E+\left(a-b\right)^{2}/4a\right]^{2}$, respectively. The blue-shaded area highlights the energy range in which the modes contained within the waveguide are non-leaky.
    }
    \label{fig:asym_disp}
\end{figure}
\begin{figure}[ht]
    \centering
    \includegraphics[width=0.5\textwidth]{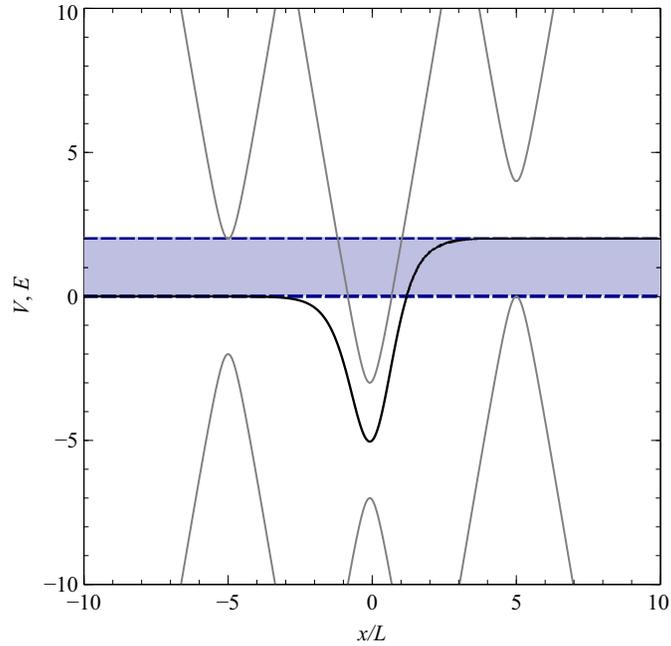}
    \caption{
(Color online) Schematic diagrams of the dispersion of a gaped Dirac material. The black lines represent the modified P{\"o}schl-Teller potential profile, of strength $a=24$ and $b=2$. The grey lines represent the charge-carrier dispersion of particles of $M=2$. The blue shaded area represent the energy range in which there are no continuum states to couple to outside of the well.
    }
    \label{fig:asym_band}
\end{figure}
We shall now consider the case of finite $b$, which represents a smooth asymmetric waveguide. Previously considered asymmetric waveguides varied abruptly on the same scale as the graphene lattice constant \cite{ping2012oscillating,he2014guided,xu2015guided,salem2016mid,xu2016guided}. We shall now consider more realistic smooth asymmetric waveguides, which fit closer to experimentally achievable potential profiles. In Fig.~\ref{fig:asym_disp}, we plot the energy spectrum for the potential defined by the parameters $a=24$ and $b=2$. The introduction of the asymmetry term $b$ reduces the number of modes at $E=0$, which are now quasi-bound modes for the massless case (Fig.~\ref{fig:asym_disp}a), since they can couple to continuum states outside of the waveguide. Naturally, for massive Dirac fermions full confinement is possible across a range of energies. In Fig.~\ref{fig:asym_band}, we show a schematic diagram of the dispersion of a gapped Dirac material, subjected to the modified P{\"o}schl-Teller potential defined by $a=24$ and $b=2$. For a particle of mass $M=2$, it can be seen that for the energy range $E=0$ to $E=2$ there are no continuum states outside of the well, therefore in that range all bound solutions will be non-leaky. The corresponding energy spectrum of confined states is shown in Fig.~\ref{fig:asym_disp}b.

\section*{Conclusions}
We have analyzed the behavior of quasi-relativistic two-dimensional particles subjected to a modified P{\"o}schl-Teller potential. Our results have direct applications to electronic waveguides in Dirac materials. For the symmetric P{\"o}schl-Teller potential, explicit forms were obtained for the eigenvalues of supercritical states. A universal expression, for any symmetric potential, was obtained for the critical potential strength required to observe the first zero-energy state. The well known results for the P{\"o}schl-Teller potential are recovered in the non-relativistic limit.

\section*{Acknowledgments}
We are grateful to Charles Downing for the critical reading of the manuscript. This work was supported by the EU H2020 RISE project CoExAN (Grant No. H2020-644076), EU FP7 ITN NOTEDEV (Grant No. FP7-607521), FP7 IRSES projects CANTOR (Grant No. FP7-612285), QOCaN (Grant No. FP7-316432), and InterNoM (Grant No. FP7-612624). R.R.H. acknowledges financial support from URCO (Project No. 09 F U 1TAY15-1TAY16) and Research Links Travel Grant by the British Council Newton Fund.

\section*{Author contributions statement}
R.R.H and M.E.P wrote the main manuscript text and R.R.H prepared the figures. All authors reviewed the manuscript.

\section*{Additional Information}
\textbf{Competing financial interests} The authors declare no competing financial interests.

\bibliography{Ref}
\end{document}